\documentclass[floats,floatfix,showpacs,amssymb,prd,twocolumn,superscriptaddress,nofootinbib]{revtex4-1}

\usepackage{amssymb,amsmath,verbatim,mathtools,needspace,enumitem,etoolbox,graphicx,physics,microtype,afterpage,bm}

\usepackage[dvipsnames, usenames]{xcolor}
\usepackage[normalem]{ulem}
\usepackage{soul}

\definecolor{linkcolor}{rgb}{0.0,0.3,0.5}
\usepackage[unicode, colorlinks=true, linkcolor=linkcolor, citecolor=linkcolor, filecolor=linkcolor,urlcolor=linkcolor, pdfusetitle]{hyperref}
\usepackage[all]{hypcap}
\usepackage[T1]{fontenc}
\usepackage[utf8]{inputenc}
\usepackage{tabularx}
\usepackage{float}
\allowdisplaybreaks
\usepackage{tensor}

\def\a{\alpha}
\def\b{\beta}

\def\d{\delta}

\def\k{\kappa}
\def\l{\lambda}
\def\m{\mu}
\def\n{\nu}

\def\p{\pi}

\def\r{\rho}
\def\s{\sigma}

\def\x{\xi}

\def\Ri{\mathcal{R}}
\def\tns{\tensor}

\newcommand{\GeV}{\; \mathrm{GeV}}

\newcommand{\be}{\begin{equation}} 
\newcommand{\ee}{\end{equation}}
\newcommand{\beq}{\begin{equation}} 
\newcommand{\eeq}{\end{equation}}
\newcommand{\bea}{\begin{equation}\begin{aligned}} 
\newcommand{\eea}{\end{aligned}\end{equation}}
\newcommand{\ba}{\begin{eqnarray}}
\newcommand{\ea}{\end{eqnarray}}

\usepackage{csquotes}
\definecolor{tclr}{RGB}{103,103,246}

\begin{document}

\title{Gravitational corrections to electroweak vacuum decay: metric vs. Palatini}

\author{Ioannis~D.~Gialamas}
\email{ioannis.gialamas@kbfi.ee}
\affiliation{Laboratory of High Energy and Computational Physics, 
National Institute of Chemical Physics and Biophysics, R{\"a}vala pst.~10, Tallinn, 10143, Estonia}

\author{Alexandros~Karam}
\email{alexandros.karam@kbfi.ee}
\affiliation{Laboratory of High Energy and Computational Physics, 
National Institute of Chemical Physics and Biophysics, R{\"a}vala pst.~10, Tallinn, 10143, Estonia}

\author{Thomas~D.~Pappas}
\email{thomas.pappas@physics.slu.cz}
\affiliation{Research Centre for Theoretical Physics and Astrophysics, Institute of Physics, Silesian University in Opava, Bezručovo nám.~13, CZ-746 01 Opava, Czech Republic}

\date{\today}

\begin{abstract} \noindent
We consider the standard Einstein-Hilbert-Higgs action where the Higgs field couples nonminimally with gravity via the term $\xi h^2 \Ri$, and investigate the stability of the electroweak vacuum in the presence of gravitational corrections in both the metric and Palatini formulations of gravity. In order to identify the differences between the two formalisms analytically, we follow a perturbative approach in which the gravitational corrections are taken into consideration via a leading order expansion in the gravitational coupling constant. Our analysis shows that in the Palatini formalism, the well-known effect of gravity suppressing the vacuum decay probability becomes milder in comparison with the metric case for any value of the nonminimal coupling $\xi$. Furthermore, we have found that in the Palatini formalism, the positivity of the gravitational corrections, which is a necessary requirement for the unitarity of the theory, entails the lower bound $\xi>-1/12$.
\end{abstract}

\maketitle

\section{Introduction} \label{sec:intro}

According to the current experimental values of the Higgs and top quark masses~\cite{ParticleDataGroup:2022pth}, the Standard Model (SM) Higgs potential, at high scales, becomes many orders of magnitude deeper than its value in the standard electroweak vacuum~\cite{Coleman:1977py, Arnold:1989cb, Sher:1988mj, Arnold:1991cv, Sher:1993mf, Casas:1994qy, Isidori:2001bm, Espinosa:2007qp, Elias-Miro:2011sqh, Degrassi:2012ry, Buttazzo:2013uya, DiLuzio:2015iua, Chigusa:2017dux, Andreassen:2017rzq}. In this way, a second minimum of the potential is realized at large field values, to which the Higgs field can decay through the nucleation of true vacuum bubbles. Nevertheless, the expected lifetime of the electroweak vacuum within the SM is much larger than the current age of the visible Universe. Therefore, the fact that bubble nucleation has not yet occurred in our past light cone is consistent with the predictions.

The issue of the gravitational effects on vacuum decay was first addressed by Coleman and De Luccia in~\cite{Coleman:1980aw}. The authors of~\cite{Isidori:2007vm} used a perturbative series in the Newton's constant to estimate this effect (see also~\cite{Branchina:2016bws} and~\cite{Markkanen:2018pdo, Devoto:2022qen} for reviews). The case of a nonminimal coupling $\xi$ between gravity and matter has been also analyzed by many authors~\cite{Rajantie:2016hkj, Czerwinska:2016fky, Salvio:2016mvj, Espinosa:2020qtq}. During inflation, gravitational effects are significant, so the stability of the electroweak vacuum has been extensively studied in this epoch~\cite{Kobakhidze:2013tn, Enqvist:2013kaa, Fairbairn:2014zia, Enqvist:2014bua, Kobakhidze:2014xda, Hook:2014uia, Herranen:2014cua, Kamada:2014ufa, Shkerin:2015exa, Kearney:2015vba, Espinosa:2015qea, Herranen:2015ima, East:2016anr, Enqvist:2016mqj, Joti:2017fwe, Ema:2017loe, Rajantie:2017ajw, Figueroa:2017slm, DeLuca:2022cus, Li:2022ugn,  Strumia:2022kez, Yin:2022fgo}. A gravitational background may suppress or enhance the rate of vacuum decay.
An enhancement can be achieved under the assumption that the bubble nucleation takes place around black holes~\cite{Gregory:2013hja, Burda:2015isa, Burda:2015yfa, Burda:2016mou, Hayashi:2020ocn, Briaud:2022few}. Thus, the existence of primordial black holes has important implications for the stability of the electroweak vacuum~\cite{Tetradis:2016vqb, Canko:2017ebb, Gorbunov:2017fhq, Mukaida:2017bgd, Gregory:2018bdt}. As a result of the high temperatures in the early Universe, classical transitions to the true vacuum are also enhanced~\cite{Tetradis:2016vqb, Canko:2017ebb}. Contrary to popular belief, in~\cite{Strumia:2022jil} it was shown that Higgs vacuum decay triggered by primordial black holes is negligibly slow due to a non-perturbative factor that suppresses the Higgs quartic coupling $\lambda$. Additionally, see~\cite{Kohri:2017ybt, Shkerin:2021zbf, Shkerin:2021rhy} where it was also argued that black hole-induced vacuum decay is negligible.  

There are two variational principles that one can apply to the gravitational action in order to obtain the Einstein equations. The metric formulation, where the connection is the usual Levi-Civita which is completely determined by the metric, and the Palatini formulation~\cite{Palatini1919, Ferraris1982}, where the metric and the connection are treated as independent variables and one has to vary the action with respect to both of them. Even though the two formulations lead to the same equations of motion for an action which is linear in the Ricci scalar $\mathcal{R}$ and contains minimally-coupled scalar field(s), the same is no longer true for more complicated actions. Modified gravity in the context of the Palatini formulation has recently received a lot of interest, especially with regard to inflation (see e.g.~\cite{Bauer:2008zj}). In the context of General Relativity, unlike in the metric approach, in the Palatini approach the metric compatibility condition arises as a dynamical consequence of the equation of motion for the connection, and not as an \emph{a priori} assumption. Furthermore, since the Palatini action involves only first derivatives of the connection, an extrinsic curvature contribution is not needed in order to cancel the generated surface terms.

In this paper, we examine how gravitational corrections affect the electroweak vacuum decay in the Palatini formalism, and compare our findings with the already existing results in the metric formulation~\cite{Coleman:1980aw, Isidori:2007vm, Branchina:2016bws, Rajantie:2016hkj, Czerwinska:2016fky, Salvio:2016mvj, Markkanen:2018pdo, Devoto:2022qen}.

The outline of the paper is as follows. In section~\ref{The_decay_rate} we review the SM vacuum decay in the absence of gravity. In section~\ref{Gravitational_corrections} we describe how the connection $\Gamma^\s_{\,\,\,\m\n}$ is differentiated in metric and Palatini formulations when a nonminimal coupling term, $f(h)\mathcal{R}$, is added in the Einstein-Hilbert-Higgs action. Subsequently, following~\cite{Isidori:2007vm, Salvio:2016mvj}, we analyze the main result of this paper, which is the lowest-order gravitational corrections to electroweak vacuum decay by means of a perturbative expansion in the gravitational coupling. Finally, we summarize and conclude in section~\ref{Conclusions}. 

\section{The vacuum decay rate}
\label{The_decay_rate}
\begin{figure}[t!]
\centering
\includegraphics[width=0.48\textwidth]{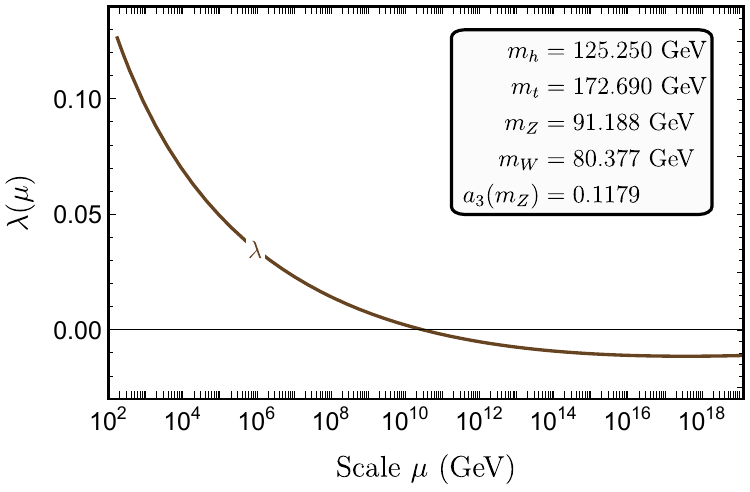}
\caption{The running of the Higgs quartic coupling  using the central values of the SM parameters~\cite{ParticleDataGroup:2022pth} shown in the inserted frame. We have used 3-loop RGEs up to the Planck scale.} 
\label{fig:coupl}
\end{figure}

Let us start by briefly reviewing the SM vacuum decay omitting gravity. The tree-level Higgs potential is given by 
\be
\label{eq:Vtree}
V(h) = \lambda(\m) \left(|H|^2 -\frac{v_h^2}{2} \right)^2 \simeq \frac{\lambda (h)}{4}h^4\,,
\ee
where $H=\left(0 \,\,\,\,\,\,\, h/\sqrt{2} \right)^T$ is the Higgs doublet, $h$ the physical Higgs field and $v_h \simeq 246.2 \GeV$ is its vacuum expectation value. The approximation
in the second equality of~\eqref{eq:Vtree} holds when considering $h \gg v_h$, as the quantum corrections to the Higgs potential can be absorbed in the running of the quartic coupling $\lambda(\m)$ at a renormalization scale $\m \sim h$. Hence, once the Higgs mass term is omitted, the remaining quartic contribution acquires instabilities for $\lambda < 0.$ Using the central values for the SM parameters~\cite{ParticleDataGroup:2022pth} (see Fig.~\ref{fig:coupl}) and 3-loop RGEs up to the Planck scale, we obtain that the Higgs running coupling becomes negative at scales larger than $2.9 \times 10^{10} \GeV$.

The RG-improved Higgs effective potential may develop a second minimum at $h\gg v_h $. Depending on the SM parameters, the electroweak vacuum is said to be stable (metastable) if the new vacuum is more (less) energetic. In the metastable case, possible tunnelling between the vacua enhances the nucleation of bubbles of true vacuum and leads to a decay of the electroweak vacuum to the true one.

In the semi-classical approximation, the tunneling rate $\Gamma$ per spacetime volume $V$ is given by~\cite{Coleman:1977py}
\be
\label{eq:Gamma}
\Gamma/V = A e^{-(S_E(h_b)-S_E(h_{\text{false}}))/\hbar} \left[1+\mathcal{O}\left( \hbar \right) \right]\,,
\ee
where $S_E(h_{\text{false}})$ is the Euclidean action of a constant solution sitting at the false vacuum and $S_E(h_b)$\footnote{From now on we omit the subscripts $``E"$ and $``b"$.} is the Euclidean action of the so-called bounce solution $h_b$, that is the solution of the Euclidean equations of motion that interpolates between the false vacuum and the opposite side of the barrier. The prefactor A is known in the flat-space case~\cite{Callan:1977pt, Isidori:2001bm}, but has yet to be computed in a curved background.

In general, the bounce solution depends only on the radial coordinate $r$ and the Euclidean equation that has to be solved reads
\be 
\label{eq:eucl_eom}
h''(r) + \frac{3}{r}h'(r)  = \frac{{\rm d }V(h)}{{\rm d} h}\,,
\ee
combined with the boundary conditions
\be
\label{eq:bndcon}
\lim\limits_{r\to \infty} h(r) = h_{\text{false}}= v_h\,,\qquad
h'(r)|_{r=0}=0\,.
\ee
The prime denotes differentiation with respect to the coordinate $r$ of a Euclidean $O(4)$-symmetric geometry. The assumption of an $O(4)$-symmetric solution is justified, as it has been proven~\cite{Coleman:1977th} that this type of solutions reproduces the minimum Euclidean action needed for a non-negligible tunnelling rate.
In the SM neglecting gravity, assuming that the effective potential for the Higgs is $\lambda h^4/4$ and that $\lambda$ has no running, it is easy to find a tree-level analytical solution of Eq.~\eqref{eq:eucl_eom} for constant negative values of the quartic coupling $\l$. This is the so-called Fubini or Lee-Weinberg bounce~\cite{Fubini:1976jm, Lee:1985uv} and reads
\be
\label{eq:h_0}
h_0(r) = \sqrt{\frac{2}{\abs{\l}}}\frac{2R}{r^2 + R^2}\,,
\ee
with $R$ being an arbitrary scale of the bounce. The arbitrariness in the bounce scale reflects the scale invariance of the tree-level field equation (under the aforementioned assumptions) which is lifted when gravitational corrections are taken into account. Then, the bounce scale R is uniquely identified with $1/\mu$, where $\mu$ is the energy scale at which the Euclidean action for the Fubini bounce
\beq\label{eq:action_S_fub}
S_0 = \frac{8\pi^2}{3 \abs{\l(\m)}}\,,
\eeq
is minimized. In the absence of gravity, this minimization occurs at $\m\sim 3.6\times 10^{17} \GeV$. In the big picture, additional contributions are incorporated in the Euclidean action, e.g.~from 1-loop corrections~\cite{Isidori:2001bm}, thermal ones~\cite{Arnold:1991cv, DelleRose:2015bpo, Salvio:2016mvj} and gravitational corrections~\cite{Coleman:1980aw, Isidori:2007vm, Branchina:2016bws, Markkanen:2018pdo, Devoto:2022qen, Rajantie:2016hkj, Czerwinska:2016fky, Salvio:2016mvj} as well. In this work, we deal only with the gravitational ones, since our aim is to compare this type of corrections in two different formulations of gravity.

\newpage
\section{Gravitational corrections}
\label{Gravitational_corrections}
As discussed in section~\ref{The_decay_rate}, the study of the metastability of the SM vacuum involves high energy scales, as the running quartic coupling $\lambda$ reaches its lowest value at scales around $\sim 10^{17} \GeV$. This scale is not so far from the Planck scale $M_{\rm Pl}\simeq 2.43\times 10^{18} \GeV$ at which, typically, the gravitational interactions become relevant, so the discussion about the gravitational corrections is a natural consequence. Despite the fact that this issue was first addressed in~\cite{Coleman:1980aw} we closely follow the careful analysis of~\cite{Isidori:2007vm, Salvio:2016mvj}.

\begin{figure}[t!]
\centering
\includegraphics[width=0.48\textwidth]{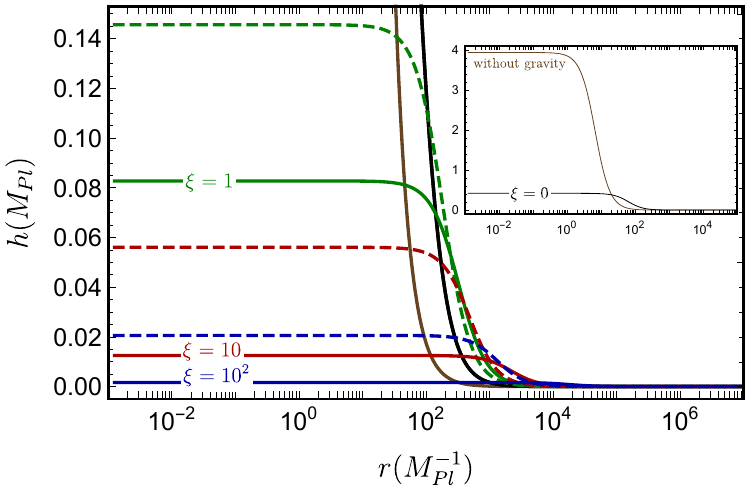}
\caption{The bounce solution~\eqref{eq:h_0} versus the radial coordinate $r$ for various values of the nonminimal coupling $\xi$. The solid lines correspond to the Palatini formulation of gravity while the dashed ones to the metric. Same color indicates same $\xi$. } 
\label{fig:hbounce}
\end{figure}

We consider a theory which contains a nonminimally coupled scalar field $h$ and a metric-independent connection $\Gamma$, specified by the Euclidean action\footnote{To obtain the Euclidean action we set the real-time coordinate $t$ to be $t=-i\tau$ and perform an analytic continuation to the metric, \textit{i.e.} $(-,+,+,+) \rightarrow (+,+,+,+).$  Our conventions for the Riemann and Ricci tensors read $\tns{\mathcal{R}}{^\a_\m_\b_\n} = \partial_\b \tns{\Gamma}{^\a_\m_\n} - \partial_\n\tns{\Gamma}{^\a_\m_\b} +\tns{\Gamma}{^\a_\s_\b}\tns{\Gamma}{^\s_\n_\m} -\tns{\Gamma}{^\a_\s_\n} \tns{\Gamma}{^\s_\b_\m}$ and $\tns{\mathcal{R}}{_\m_\n} = \tns{\mathcal{R}}{^\a_\m_\a_\n}$. The Ricci scalar is obtained by the contraction $\mathcal{R} = g^{\m\n}\mathcal{R}_{\m\n}$.}\\

\newpage

\ba\label{eq:EHH_Palatini}
  S &=& \int\dd^4 x \sqrt{g} \bigg[ -\frac{g^{\m\n}\mathcal{R}_{\m\n}(\Gamma)}{2\k}  - \frac{g^{\m\n}\mathcal{R}_{\m\n}(\Gamma)}{2} f(h)  \nonumber
  \\ && +  \frac{1}{2} g^{\m\n}\partial_\m h \partial_\n h + V(h) \bigg] \, ,
\ea
with $\k = 1/M_{\rm Pl}^2$. Variation with respect to the (torsionless) connection gives
\be \label{eq:Gamma_Pal}
\Gamma^\s_{\,\,\,\m\n}= \overline{\Gamma}^\s_{\,\,\,\m\n} + \d_F \left(\d^\s_{\,\,\,\m}\partial_\n \omega(h) +\d^\s_{\,\,\,\n}\partial_\m \omega(h) -g_{\m\n}\partial^\s \omega(h) \right)\,,
\ee
where
\be \label{eq:omega}
\omega(h) = \ln \sqrt{1+  \k f(h)}\,.
\ee
$\overline{\Gamma}^\s_{\,\,\,\m\n}$ is the  Levi-Civita connection and $\d_F =1$ in the Palatini formulation, while $\d_F =0$ holds for the metric one.

Our objective is to compute the gravitational corrections to the flat-space result for the electroweak-vacuum decay, in the context of the Palatini formalism. So, our starting point is the Euclidean Einstein-Hilbert-Higgs action 
\be \label{eq:EHH_action}
  S = \int\dd^4 x \sqrt{g} \left[ -\frac{\mathcal{R}}{2\k}  - \frac{\mathcal{R}}{2} f(h)  +  \frac{1}{2}  g^{\m\n}\partial_\m h \partial_\n h + V(h) \right]\,,
\ee
in which we allow a renormalizable~\cite{Toms:1982af, Espinosa:2007qp, Herranen:2014cua} nonminimal coupling, between gravity and Higgs of the form $f(h) = \xi h^2$, with $\xi$ being a dimensionless parameter\footnote{Notice that in a more in-depth analysis, one has to deal with a varying $\xi$, as it is subject to quantum corrections. However, in this work, we assume that its running is suppressed and can be safely ignored. We leave the calculation of the running of $\xi$ in the Palatini case for future work.}. The Euclidean equations of motion are
\ba\label{eq:Eucl_EOM1}
\mathcal{R}_{\m\n}-\frac{1}{2}g_{\m\n}\mathcal{R} &=& \k T_{\m\n}^{\text{eff}}\,,
\\ \overline{\nabla}_\m \overline{\nabla}^\m h +\xi h \mathcal{R} &=& \frac{{\rm d }V(h)}{{\rm d} h}\,, \label{eq:Eucl_EOM2}
\ea
where the effective energy-momentum tensor is defined as
\begin{widetext}
\be\label{eq:Tmneff}
T_{\m\n}^{\text{eff}} = \frac{\partial_\m h \partial_\n h - g_{\m\n}\left(\frac{1}{2}\partial_\l h \partial^\l h + V(h) \right) +\xi(1-\d_F)\left(\overline{\nabla}_\m \overline{\nabla}_\n h^2 -g_{\m\n} \overline{\nabla}_\l \overline{\nabla}^\l h^2 \right)}{1+\k\xi h^2}\,,
\ee
\end{widetext}
and the $ \overline{\nabla}_\m$ is the covariant derivative constructed from the Levi-Civita connection.

One can go to the Einstein frame by performing a Weyl rescaling of the form $g_{\m\n} \rightarrow \left(1+\k \xi h^2 \right)g_{\m\n}$. In this way the nonminimal coupling $\xi$ decouples from the Ricci scalar and its effect is transferred to the scalar potential. A field redefinition is also needed in order to get a canonical scalar field. In any case, the Euclidean action in both the Einstein and Jordan frames is the same, as the Weyl rescaling of the metric transforms the field without affecting the action. Thus, one is free to choose whichever frame is the most convenient one to perform the computations in. For the model we consider here, the optimal choice is the Jordan frame action of Eq.~\eqref{eq:EHH_action} since the potential is much simpler and there is no need for a field redefinition.

As Eq.~\eqref{eq:Eucl_EOM2} does not include explicitly the ``formalism'' term $\d_F$, one could claim that this equation is invariant in the two formalisms. This is not true because the Ricci scalar $\mathcal{R}$ is not the same in the two formalisms as we will see below.  Both Eqs.~\eqref{eq:Eucl_EOM1} and~\eqref{eq:Eucl_EOM2} reduce to their known metric case's form once we set $\d_F=0$. 

\begin{figure*}[t!]
\centering
\includegraphics[width=0.7\textwidth]{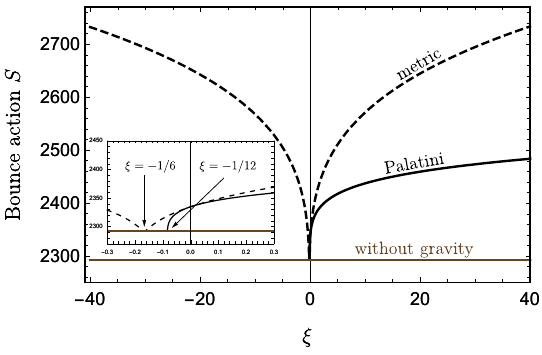}
\caption{The minimum values of the bounce action~\eqref{eq:correction} (black lines) for the Palatini (solid) and metric (dashed) formalism versus the nonminimal coupling $\xi$. The brown line indicates the minimum value of the $S_0$ action, i.e. the case when gravity is absent.} 
\label{fig:sVSxi}
\end{figure*}

As discussed in Sec.~\ref{The_decay_rate}, in a flat spacetime the lowest value of the action is achieved by means of an $O(4)$-symmetric bounce solution of the equations of motion. In our case, the inclusion of gravity possibly alters the type of the solution, but we will follow the usual prescription assuming a Euclidean $O(4)$-symmetric geometry~\cite{Coleman:1980aw}
\be \label{eq:O(4)_metric}
\rm d s^2 =\rm d r^2 + \r^2(r) \rm d \Omega_3^2 \,,
\ee
as in the case of a non-gravitational background. Even so, there is a possibility that a bounce solution with a smaller action and no $O(4)$ symmetry might exist.
In~\eqref{eq:O(4)_metric}, $\rm d \Omega_3^2 = \rm d \phi^2 + \sin^2\phi\, \left(\rm d \theta^2 +\sin^2\theta \, \rm d \varphi^2 \right)$  is the volume element of the unit $3$-sphere and $\r$ gives the radius of curvature of each $3$-sphere. In this background, the action simplifies to
\be \label{eq:EHH_action_2}
  S = 2\pi^2\int\rm d r \r^3  \left[ -\frac{\mathcal{R}}{2\k}  - \frac{\mathcal{R}}{2} f(h)  +  \frac{h'^2}{2} + V(h) \right] \, .
\ee
 The Ricci scalar is given by
\ba \label{eq:Ricci_Pal}
 \mathcal{R}&=&\frac{- 6 \left( \r^2 \r'' +\r \r'^2 -\r \right)}{\r^3} \nonumber
 \\ &&- 6\d_F \left(\omega''(h) + \omega'^2(h) +3 \omega'(h) \frac{\r'}{\r} \right)\,.
\ea
The second term in Eq.~\eqref{eq:Ricci_Pal} that appears only in the Palatini case is of the same order $(\mathcal{O}(\k))$, in a $\k$-expansion,  with the first term.
In the $O(4)$-symmetric geometry~\eqref{eq:O(4)_metric},
the $rr$ component of the Einstein equation~\eqref{eq:Eucl_EOM1} is
\ba \label{eq:Eeq_rr}
\r'^2 &=& 1 +\frac{\k \r^2}{3\left(1+\k \xi h^2 \right)} \left(\frac{h'^2}{2}-V(h)-6\frac{\r'}{\r}\xi h h' \right) \nonumber
\\ &&-\d_F \omega'^2(h) \r^2\,,
\ea
while the Klein-Gordon equation~\eqref{eq:Eucl_EOM2} takes the form
\be \label{eq:Klein-Gordon}
h'' + 3\frac{\r'}{\r}h'  = \frac{{\rm d} V(h)}{\rm d h}-\xi h \mathcal{R}\,.
\ee
The new boundary conditions are those of Eq.~\eqref{eq:bndcon} along with the gravity-imposed conditions $\rho(0) = 0$ and $\rho(\infty) = r$. Adding these conditions, ensures that $r=0$ is at the centre of the bounce and that a Minkowski-like background is laid out in the false vacuum.

The calculation of the bounce action requires the solution of the coupled system of field equations~\eqref{eq:Eeq_rr}-\eqref{eq:Klein-Gordon}, and this can only be achieved by means of numerical integration. However, in order to have a better understanding of the results, we will follow instead the very accurate approximate analytic prescription proposed in~\cite{Isidori:2007vm, Salvio:2016mvj} which, as we will show, is sufficient to encapsulate the differences emerging between the two formulations. Following the perturbative method applied in~\cite{Isidori:2007vm, Salvio:2016mvj}, we perform a leading-order expansion in the gravitational coupling $\k$:
\ba\label{eq:expansion}
h(r) &=& h_0(r) +\k h_1(r) + \mathcal{O}(\k^2)\,, \qquad \nonumber
\\ \r(r) &=& r +\k \r_1(r)  +\k^2 \r_2(r) + \mathcal{O}(\k^3)\,,
\ea
where only the $\mathcal{O}(\k^0)$ terms survive in the absence of gravity. The metric function $\r$ is expanded to order $\mathcal{O}(\k^2)$, because of the $1/\k$ coefficient in the Ricci scalar.  In any case, the approximation of expanding the functions $h(r)$ and $\r(r)$ in powers of $\k$ is absolutely valid, since the higher-order terms are Planck suppressed. A more precise numerical analysis confirms our claim, as the bounce action which is calculated below analytically is in good agreement with its numerical value as the error is about $2\%$ for $\xi \sim 40$ (in both formulations), being decreased for smaller values of $\xi$.  Hence, a fully numerical treatment of the problem does not significantly affect the conclusions drawn with our approximate-analytic approach.

The action~\eqref{eq:EHH_action_2} at  $\mathcal{O}(\k)$ is
\beq\label{eq:action_exp}
S= S_0 +S_\k + \mathcal{O}(\k^2)\,,
\eeq
with
\beq\label{eq:action_S_0}
S_0 = 2\pi^2 \int {\rm d} r\, r^3 \left(\frac{h_0'^2}{2}+V(h_0) \right)\,,
\eeq
and
\ba\label{eq:action_S_k}
S_\k &=& 6\pi^2 \k \int {\rm d} r \bigg[ r^2 \r_1  \left(\frac{h_0'^2}{2}+V(h_0) \right) \nonumber
\\ &&+ \left(r \r_1'^2 +2\r_1 \r_1' +2r\r_1 \r_1''  \right)  \nonumber
\\ &&+  r f(h_0) \left(r \r_1'' +2 \r_1' \right) - \d_F \x r^3 f(h_0) h_0'^2  \bigg]\,.
\ea
There are additional total-derivative (t.d.) terms in the action~\eqref{eq:action_exp} that vanish upon integration. These terms read
\ba
S^{\text{t.d.}} &=& 6\pi^2 \int {\rm d} r \left[ (r^2 \r_1')' + \d_F\x (r^3h_0h_0')'\right] \nonumber
\\ \ && + 6\pi^2 \k \int {\rm d} r  \Bigg[\left(\frac{r^3h_0'h_1}{3}\right)' + (r^2\r_2')' \nonumber
\\ && + \d_F\x (r^3h_0'h_1+r^3h_0h_1' +3r^2\r_1h_0h_0')' \Bigg].
\ea
The term $(r^3 h_0' h_1/3)' $ appears after using the Klein-Gordon equation~\eqref{eq:Klein-Gordon} in flat space, i.e. Eq.~\eqref{eq:eucl_eom}.

Further simplification of the gravitational action~\eqref{eq:action_S_k} may occur upon applying the rescaling $\r_1 \rightarrow s \r_1$ and requiring that $\rm d S_k / \rm d s |_{s=1}=0$. The gravitational action at $\mathcal{O}(\kappa)$ then acquires the handy form
\ba\label{eq:action_S_k_final}
S_\k = 6\pi^2 \k \int {\rm d} r\left[r \r_1'^2 -\d_F \xi^2 r^3 h_0^2 h_0'^2 \right]\,.
\ea
The final step is to expand both sides of Eq.~\eqref{eq:Eeq_rr} at leading order in $\k$ to obtain that 
\be
\label{eq:r_1'}
\r_1'(r) = \frac{r^2}{6}\left(\frac{h_0'^2}{2}-V(h_0) -\frac{6\xi}{r} h_0 h_0' \right)\,,
\ee
with $h_0$ being the Fubini bounce given by Eq.~\eqref{eq:h_0}. 
Notice here that~\eqref{eq:r_1'} is $\d_F$-independent as the $-\d_F \omega'^2(h) \r^2$ term in the right-hand-side of~\eqref{eq:Eeq_rr} is $\mathcal{O}(\k^2)$. Substituting Eqs.~\eqref{eq:h_0} and \eqref{eq:r_1'} in Eqs.~\eqref{eq:action_S_0} and~\eqref{eq:action_S_k_final} we obtain
\beq\label{eq:correction}
S = \frac{8\pi^2}{3 \abs{\l(\m)}} + \frac{32 \p^2 \k\left( 1+12\xi +36\xi^2(1-\d_F) \right)}{45 R^2 \l^2(\m)}\,.
\eeq
The first term corresponds to the bounce action in the absence of gravity and is of course the same in both metric and Palatini cases. The second term encapsulates the gravitational correction to the action and clearly exhibits a dependence on the formalism. This correction vanishes for $\xi=-1/6$ in the metric case ($\d_F=0$), and for $\xi=-1/12$ in the Palatini case ($\d_F=1$). For $\xi=0$, the gravitational correction becomes independent of the formalism since in the minimally-coupled case the two formalisms are equivalent.
\begin{figure}[t!]
\centering
\includegraphics[width=0.48\textwidth]{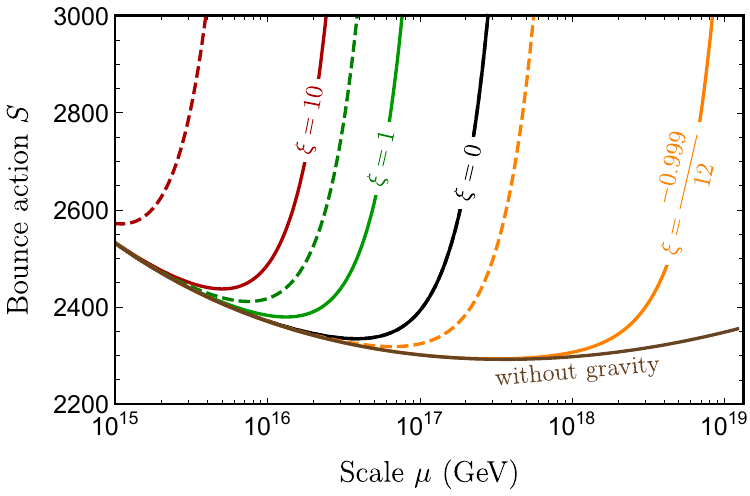}
\includegraphics[width=0.48\textwidth]{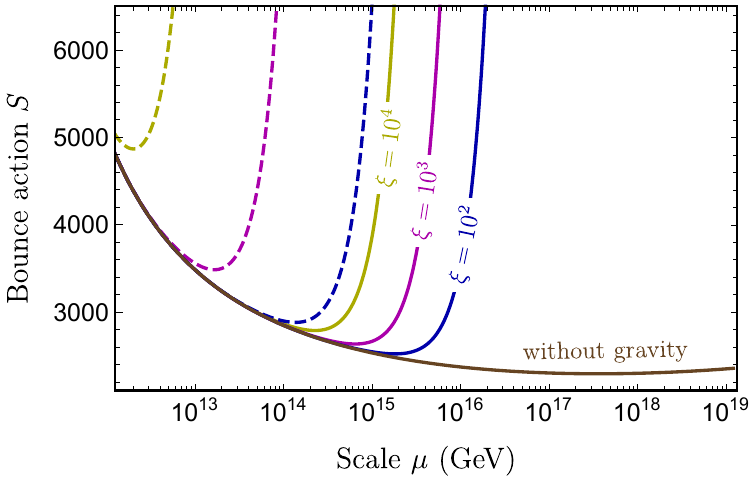}
\caption{The bounce action~\eqref{eq:correction} for the Palatini (solid) and metric (dashed) formalism versus the scale $\mu$ for various values of the nonminimal coupling $\xi$. The brown line indicates the $S_0$ action, i.e. the case when gravity is absent. Same color indicates same $\xi$.} 
\label{fig:sVSmu}
\end{figure}
In Fig.~\ref{fig:hbounce} the bounce solution~\eqref{eq:h_0} is depicted for various values of the nonminimal coupling $\xi$. The $\xi$-dependence on $h_0$ arises from the $\xi$-dependent scale $\mu$ at which the action~\eqref{eq:correction} is minimized. In this way the true bounce $h(r)$ is correctly approximated by the Fubini bounce $h_0(r)$ with $1/R=\mu$ being the value that minimizes the gravitational corrected action~\eqref{eq:correction} (see~\cite{Isidori:2007vm}). From this figure, we can see that the value of the bounce solution is reduced in the context of Palatini gravity. Fig.~\ref{fig:sVSxi} illustrates the bounce action with (black lines) and without (brown line) gravity in metric and Palatini formalisms. Clearly, for any nonzero value of $\xi$, the Palatini formalism yields a bounce action that is suppressed w.r.t. the metric case, and as a consequence, EW vacuum decay in the presence of gravity becomes more probable when the Palatini formalism is utilized. For values of $\xi<-1/12$, the Palatini gravitational correction becomes negative and thus the vacuum decay probability can become $\geq 1$ and as a consequence, unitarity is not preserved. In the metric case both negative and positive values of $\xi$ are allowed since the correction is always positive as it is $\sim (1+6\xi)^2$~\cite{Salvio:2016mvj}. Finally, in Fig.~\ref{fig:sVSmu} the bounce action is displayed for different values of $\xi$ in a wide range of scales. Each minimum of these curves coincides with the minimum bounce action values of Fig.~\ref{fig:sVSxi}. For the same $\xi$ the bounce action is extremized at higher scales in the Palatini case. The ratio of the extremization scales $\m_\text{extr}^{P/m}$ in the two formulations becomes larger as the absolute value of the parameter $\xi$ increases, e.g. $\m_\text{extr}^P/\m_\text{extr}^m |_{\xi=1} \sim \mathcal{O}(1)$, while  $\m_\text{extr}^P/\m_\text{extr}^m  |_{\xi=10^4} \sim \mathcal{O}(10^2)$.\\

\section{Conclusions and discussion}
\label{Conclusions}
In this work, we have considered the gravitational corrections to electroweak vacuum decay in the context of the Palatini formulation of gravity, and we have compared our findings with the already-known results in the metric formulation. Although in the context of General Relativity, the two formulations are equivalent, in the presence of fields that are coupled in a nonminimal manner to gravity this no longer holds. 

We saw that gravity suppresses the electroweak vacuum decay less in the Palatini case than in the usual metric approach but still, it improves on the vacuum metastability problem. Also, the linear $\xi$-dependence of the gravitational correction indicates that the Palatini nonminimal coupling has to be larger than the critical value $-1/12$ in order to avoid unitarity issues. Also, for $\xi=-1/12$ the gravitational corrections disappear in the Palatini case, while in the metric case this happens when the parameter $\xi$ takes the so-called conformal value, that is $\xi=-1/6$. The physics behind this particular value of the parameter $\xi$ in the Palatini formalism will be the subject of future work.

The differences between the metric and Palatini formulations of gravity are most commonly studied in the context of inflation. For the model at hand~\cite{Bezrukov:2007ep, Bauer:2008zj}, the phenomenological distinction can be seen in the predicted value for the tensor-to-scalar ratio $r$. The metric version of the theory predicts $r_{\rm metric} \sim \mathcal{O} (10^{-3})$ while the Palatini version predicts $r_{\rm Pal} \sim \mathcal{O}(10^{-12})$ for $\lambda = 0.1$ (and higher values of $r_{\rm Pal}$ for smaller $\lambda$, but always $r_{\rm Pal} < r_{\rm metric}$). Future planned experiments such as PICO~\cite{Hanany:2019lle} will have a sensitivity of $\delta_r \sim \mathcal{O}(10^{-4})$ and could thus falsify the metric version of the theory (along with the Starobinsky model~\cite{Starobinsky1980}). If that turns out to be the case, then one could claim that the Palatini formalism is on a better footing than the metric one. Furthermore, it would be interesting to study Higgs vacuum decay during inflation in the two formalisms, but this will also be the topic of future work.

\vspace{.5 cm}
\paragraph*{ Acknowledgments.}
IDG likes to thank M.~Raidal and N.~Tetradis for useful discussions. The work of IDG was supported by the Estonian Research Council grant SJD18. AK was supported by the Estonian Research Council grant PSG761, MOBTT86, and by the EU through the European Regional Development Fund CoE program TK133 ``The Dark Side of the Universe". TDP acknowledges the support of the Research Centre for Theoretical Physics and Astrophysics of the Institute of Physics at the Silesian University in Opava.

\bibliography{main}

\end{document}